\documentclass[10pt]{article}
\usepackage{amsfonts, amssymb, amsthm,amsmath}

\renewcommand{\theequation} {\arabic{section}.\arabic{equation}}

\newcounter{orange}
\renewcommand{\theorange}{\alph{orange}}

\newcommand{\be}{\begin{equation}}
\newcommand{\ee}{\end{equation}}
\newcommand{\bea}{\begin{eqnarray}}
\newcommand{\eea}{\end{eqnarray}}

\begin{document}


\begin{center}
{\LARGE{\bf{Notes on thermodynamics in special relativity}}}
\end{center}

\begin{center}
\vskip0.25cm

{\bf M. Przanowski \footnote{E-mail address:przan@fis.cinvestav.mx} and J.  Tosiek \footnote{E-mail address:tosiek@p.lodz.pl}}  \\

\vskip0.25cm
{\em  Institute of Physics}\\ {\em Technical University of Lodz}\\ {\em ul. Wolczanska 219, 90-924 Lodz}\\
{\em Poland}
\end{center}
\vskip0.15cm

\vskip0.25cm\centerline{\today}
\begin{center}
\begin{abstract}
Foundations of thermodynamics in special theory of relativity are considered. We argue that from the phenomenological point of view the correct relativistic transformations of heat and absolute temperature are given by the formulas proposed by H. Ott, H. Arzeli\`{e}s and C. M\o ller. It is shown that the same transformation rules can be also found from the relativistic Gibbs distribution for ideal gas. This distribution has been recently verified by the computer simulations. Phenomenological and statistical thermometers in relativistic thermodynamics  are analyzed.
\end{abstract}

\end{center}

\section{Introduction}

\label{section1}

The present paper is an extended and improved version of the previous work of one of us (MP) \cite{mp} and is devoted to the foundations of relativistic thermodynamics in special theory  of relativity. Referring the reader to the next section for details we note here that there is a long standing problem concerning the relativistic transformation rules for heat and temperature. First, A. Einstein \cite{ae}, M. Planck \cite{max}, K. von Mosengeil \cite{kmo}, W. Pauli \cite{wp}, M. von Laue \cite{ml} and many others found  transformation formulas (\ref{2.18}) and  (\ref{2.28}). In 1952 A. Einstein  changed his primary opinion in favour of  new formulas (\ref{2.17}) and (\ref{2.27}) \cite{chl} which have been also found later on
by H. Ott \cite{hot}, H. Arzeli\`{e}s \cite{har} and C. M\o ller \cite{cmo}--\cite{cm3}. (It is worthwhile to point out that in the first edition of M\o ller's book in 1952 the `classical' transformation rules (\ref{2.18}) and  (\ref{2.28}) were considered to be correct and in the second edition from 1972 \cite{cm3} the things have been changed and the formulas (\ref{2.17}) and (\ref{2.27}) have been given as the correct ones). 
Note that perhaps L. D. Landau and E. M. Lifshitz were the first authors who mentioned the transformation formula for $T$ equivalent to (\ref{2.27}) in the Russian edition of their famous monograph on statistical physics in 1951 (see Eq. (27.4) in \cite{ldl}).

But it is not the all story. P. T. Landsberg and his collaborators  in some works \cite{ptl1}--\cite{ptl4} argue that heat and temperature are Lorentz invariants i.e. 
(\ref{2.19}) and  (\ref{2.29}) hold true and then in some other works \cite{ptl5}--\cite{ssc} they claim that there does not exist any universal relativistic transformation of temperature. Recently this point of view has been strongly supported in \cite{gls1}, \cite{gls2}.
The idea that heat and temperature are Lorentz invariants has been also analyzed by N. G. van Kampen \cite{ngvk}.

The above very brief review of development of opinions on the transformation rules for heat and temperature in relativistic thermodynamics does not sound optimistically. It seems that the eventual relativistic thermodynamics depends very much on  accepted conventions and that there does not exist any natural system of conventions  which leads to a correct construction of relativistic thermodynamics. However, in our opinion it is not so. A strong confirmation of this opinion we have found in distinguished papers by  C. M\o ller \cite{cmo}--\cite{cm2} and, recently in a nice work of M. Requardt \cite{mr}. In the mentioned works it is argued that there exists a natural definition of thermodynamic work and then, by the first law of thermodynamics, the natural definition of heat. Moreover, from the second law of thermodynamics one arrives at the natural definition of  absolute temperature. This temperature can be experimentally determined with the use of an appropriate cyclic thermodynamic engine. Having defined heat and temperature it is shown that the respective transformation rules are given by  (\ref{2.17}) and (\ref{2.27}). In Sec. \ref{section2}  we analyze all these questions not only for ideal liquid but also for an arbitrary continuous medium and finally we arrive at the conclusion that the correct transformation formulas for heat and absolute temperature in relativistic thermodynamics indeed are given by (\ref{2.17}) and (\ref{2.27}). 

It is expected that, analogously as in nonrelativistic thermodynamics,  deeper insight into the problems of phenomenological relativistic thermodynamics can be acquired when the respective statistical thermodynamics is applied. But here we are in trouble since, as yet, there does not exist a general relativistic statistical thermodynamics.

In 1911 F. J\"{u}ttner \cite{fj} found the relativistic generalization of the Maxwell distribution to ideal gas. The J\"{u}ttner  distribution was generalized to the case of moving ideal gas and the Bose--Einstein or Fermi--Dirac ideal gases \cite{rkp1}, \cite{rkp2}.
At first the J\"{u}ttner  distribution was commonly accepted \cite{wp}, \cite{rkp1}--\cite{srg}, but then several objections against this distribution have been raised \cite{lph1}--\cite{jd}.

Nevertheless, we meet a light at the end of this tunnel provided by recent computer simulations \cite{dc}--\cite{pea}. Numerical results of the simulations are in a perfect agreement with the $1$--D J\"{u}ttner  distribution and its generalization to the moving gas \cite{dc}, \cite{jd2} or with the $2$--D J\"{u}ttner  distribution \cite{cr}--\cite{mh2} and  its generalization to the moving  frame \cite{mh1}, \cite{mh2}. In $3$--D case Monte Carlo simulations also confirmed the J\"{u}ttner  function \cite{pea}.

Therefore assuming that the $3$--D J\"{u}ttner  distribution 
and its generalization to a moving system 
are the correct relativistic equilibrium distributions we define the relativistic Gibbs distribution for ideal gas with respect to any inertial frame (\ref{3.1}). Then in Sec. \ref{section3}  we use the well known in nonrelativistic statistical physics machinery \cite{ldl} leading from the Gibbs distribution to phenomenological thermodynamics and we prove that from this point of view the correct transformations of heat and absolute temperature are given again by (\ref{2.17}) and (\ref{2.27}) respectively.

In Sec. 4 some useful identities in phenomenological relativistic thermodynamics are found and the methods of measuring the absolute temperature are proposed. Section 5 is devoted to searching for a natural definition of the {\bf statistical thermometer}. In fact one can find many `natural' statistical thermometers. One of them can be constructed with the use of the generalized principle of equipartition of energy (\ref{5.11a}).

However, if one searches for some parameter which independently of the reference frame characterizes  systems being in 
thermal equilibrium then this parameter is certainly the {\bf empirical temperature} $T_0$ and this is what has been pointed out in \cite{dc}. From this point of view  the natural statistical thermometer is the one defined by (\ref{5.21}) (see also \cite{dc}).

We can also define a statistical thermometer which measures the absolute temperature $T_0$ of the ideal gas at rest (see (\ref{5.21})) which is equivalent to the empirical temperature or another one which measures the absolute temperature $T$ (see (\ref{5.20})). Many other possibilities are also acceptable (e.g. (\ref{5.22})).

In conclusion we find that there are two notions of relativistic temperature. One of them follows from the zeroth law of relativistic thermodynamics and this is the {\bf empirical temperature}, which is a relativistic scalar and which can be identified with the absolute temperature $T_0$ of a thermodynamic system at rest. 
It seems that the analogous point of view concerning $T_0$ is represented by A. Staruszkiewicz, who calls $T_0$ a 'relative scalar', since it is measured in the proper frame of a thermodynamic system \cite{stz}.

The second one is a consequence of the second law of relativistic thermodynamics and this is the {\bf absolute temperature} $T$ of the transformation law (\ref{2.27}). Although in nonrelativistic thermodynamics one also meets these two notions of temperature, this is only relativistic thermodynamics which shows explicitly the deep difference between them.

We use the Einstein summation convention. The  symbol $p$ denotes either a momentum or a pressure but from the context its meaning  is always clear. The subindex `$0$' corresponds to the proper frame $K_0$ of the thermodynamic system.


\section{Phenomenological relativistic thermodynamics}

\label{section2}

\setcounter{equation}{0}

Consider a  continuous medium in  thermodynamic equilibrium. We choose a bounded domain of the medium of the volume $V_0$ with respect to the proper inertial frame $K_0.$ We assume also that the medium moves with a constant velocity  $\vec{ v}$ with respect to the laboratory inertial frame $K.$ The space axes of $K_0$ and $K$ are assumed to be mutually parallel. The spacetime metric is $\eta_{ij}=\eta^{ij}={\rm diag}(-,-,-,+).$

Then the coefficients $\Lambda^j_k, \; j,k=1,2,3,4$ of the Lorentz transformation $x^j=\Lambda^j_k x_0^k$ are given by the formulas
\be
\label{2.1}
\Lambda^{\mu}_{\nu}=\delta^{\mu}_{\nu} - \frac{v^{\mu}v^{\nu}}{v^2}\big(1- \gamma(v) \big) \;\;, \;\; \Lambda^{\mu}_4=\frac{v^{\mu}}{c}\gamma(v)\;\;,\;\; \Lambda^{4}_{\mu}=\frac{v^{\mu}}{c}\gamma(v)\;\;,\;\; \Lambda^4_4=\gamma(v)\;\;,\;\; \mu,\nu=1,2,3,
\ee
 where $\gamma(v):=\frac{1}{\sqrt{1-\frac{v^2}{c^2}}}.$

The energy--momentum tensor $T^{jk}$ for a continuous medium can be written in the form \cite{cm3}, \cite{ldl2}, \cite{wei}
\be
\label{2.2}
T_{jk}=\varepsilon_0 u_j u_k + \tau_{jk}=T_{kj},   
\ee
where $\varepsilon_0$ is the energy density in the proper inertial frame   $K_0,$  $u^j=\big( \frac{v^{\mu}}{c}\gamma(v), \gamma(v)\big)$ is the $4$--vector of velocity of $K_0$ relative to $K,$ and $\tau^{jk}$ is determined by the stress tensor $\tau_0^{\;\mu \nu}$ as follows
\be
\label{2.3}
\tau^{jk}= \Lambda^j_l \Lambda^k_m \tau_0^{\;l m}, \;\; \tau_0^{\;4j}= \tau_0^{\;j4}=0 \Rightarrow \tau^{jk}= \Lambda^j_{\mu} \Lambda^k_{\nu} \tau_0^{\;\mu \nu}.
\ee
We put $\tau^{4j}_0=\tau^{j4}_0=0$ because for a system in the rest the stream of strength $\tau^{4 \mu}_0=0$.
Moreover, $T_{44}=\varepsilon_0$ so 
 that $\tau^{44}_0=0$.

Inserting (\ref{2.1}) into (\ref{2.3}) and then (\ref{2.2}) one quickly finds the total energy $E$ and momentum $P^{\mu}$
\setcounter{orange}{1}
\renewcommand{\theequation} {\arabic{section}.\arabic{equation}\theorange}
\be
\label{2.4a}
E=\int_V T^{44}dx^1dx^2dx^3= \gamma(v)\left( E_0 + \frac{v^{\nu}v^{\rho}}{c^2}{\cal T}_{0 \;\nu \rho}\right),
\ee
\addtocounter{orange}{1}
\addtocounter{equation}{-1}
\be
\label{2.4b}
P^{\mu}=\frac{1}{c}\int_{V} T^{\mu 4} dx^1 dx^2 dx^3= \frac{v^{\mu}}{c^2}\gamma(v)E_0 + \frac{v^{\nu}}{c^2}\left[ \big(\gamma(v)-1 \big) \frac{v^{\mu}v^{\rho}}{v^2}-\eta^{\mu \rho}\right]{\cal T}_{0 \;\nu \rho}
\ee
\renewcommand{\theequation} {\arabic{section}.\arabic{equation}}
where ${\cal T}_{0 \;\nu \rho}:= \int_{V_0}\tau_{0 \;\nu \rho}\; dx^1_0 dx^2_0 dx^3_0.$

As yet we have not used the conservation law $\frac{\partial T^{jk}}{\partial x^k}=0.$ Since our thermodynamic system is in equilibrium this law in the proper  frame $K_0$ takes the form
\be
\label{2.5}
\frac{\partial \tau_0^{\;\mu \nu}}{\partial x^{\nu} }=0.
\ee
With the use of (\ref{2.5}) one gets \cite{ldl2}
\be
\label{2.6}
{\cal T}_0^{\;\mu \nu}= \int_{V_0}\tau_0^{\;\mu \nu}dx^1_0 dx^2_0 dx^3_0= \int_{V_0} \frac{\partial \big(\tau_0^{\;\mu \rho}\,x_0^{\nu}\big)}{\partial x_0^{\rho}}
dx^1_0 dx^2_0 dx^3_0= - \oint_{\partial V_0} \tau^{\;\,\mu}_{0 \,\rho}\, x^{\nu}_0 \,n^{\rho}_0 \,d \Sigma_0,
\ee
where $n^{\rho}_0$ is the $\rho$'s component of the unit outward normal vector to the boundary $\partial V_0$ of $V_0$ and $d \Sigma_0$ is the surface element of $\partial V_0$.

By symmetrization of  (\ref{2.6}) we finally get
\be
\label{2.7}
{\cal T}_{0\;\mu \nu}= - \frac{1}{2} \oint_{\partial V_0} \Big(\tau_{0 \;\mu \rho} \, x_{0 \;\nu}+ \tau_{0 \;\nu \rho} \, x_{0 \;\mu}\Big) n^{\rho}_0 \,d \Sigma_0.
\ee
In particular in the case of ideal liquid one has
\be
\label{2.8}
T_{jk}=(p_0 + \varepsilon_0)u_j u_k - p_0 \eta_{jk} \;\;,\;\; \tau_{0 \;\mu \nu}=-p_0 \eta_{\mu \nu}\;\;,\;\; {\cal T}_{0 \;\mu \nu}=-p_0 V_0 \eta_{\mu \nu}
\ee
where $p_0$ stands for the pressure. Tensor $T_{jk}$ in (\ref{2.8}) can be rewritten in the form of (\ref{2.2}) as follows
\[
T_{jk}= \epsilon_{0} u_j u_k + p_0 (u_j u_k - \eta_{jk})
\]
(compare \cite{ldl2}, \cite{wei}).

Substituting (\ref{2.8}) into (\ref{2.4a}) and (\ref{2.4b}) we obtain
\setcounter{orange}{1}
\renewcommand{\theequation} {\arabic{section}.\arabic{equation}\theorange}
\be
\label{2.9a}
E= \gamma(v)\left(E_0 + \frac{v^2}{c^2}p_0\, V_0\right),
\ee
\addtocounter{orange}{1}
\addtocounter{equation}{-1}
\be
\label{2.9b}
P^{\mu}= \gamma(v)\Big(E_0 + p_0\, V_0 \Big)\frac{v^{\mu}}{c^2}.
\ee
\renewcommand{\theequation} {\arabic{section}.\arabic{equation}}
Remembering that pressure is a Lorentz invariant i.e. $p=p_0$ and the volume $V_0=V \gamma(v)$ we can rewrite (\ref{2.9a}) as follows
\be
\label{2.10}
E + p V= \gamma(v)\big(E_0 +p_0 V_0\big).
\ee
From (\ref{2.9b}) and (\ref{2.10}) with (\ref{2.1}) one concludes that $\left( P^1,P^2,P^3, \frac{E +pV}{c}\right)$ constitutes a $4$--vector. This crucial point has been considered by many authors \cite{wp}, \cite{cmo}, \cite{cm2}, \cite{mr}, \cite{rkp1}, \cite{rkp2}, \cite{as}, \cite{chn}.

We assume that the first law of thermodynamics for our thermodynamic system in its proper frame $K_0$ has the usual form
\be
\label{2.11}
dE_0=\delta Q_0 +\delta L_0,
\ee 
where $\delta Q_0$ is the amount of heat entering  the system and $\delta L_0$ denotes the thermodynamic work done over the system. For  reversible processes one has
\be
\label{2.12}
\delta Q_0=T_0 dS_0\;\;, \;\; \delta L_0= - \int_{\partial V_0} \tau_{0 \;\mu \nu } \,n^{\nu}_0 \,dx_0^{\mu} \,d \Sigma_0
\ee
with $T_0$ being the absolute temperature and $S_0$ the entropy of the system. For example, in the case of ideal liquid, from (\ref{2.8}) and (\ref{2.12}) we obtain
\be
\label{2.13}
\delta L_0= p_0 \int_{\partial V_0} \eta_{\mu \nu } \,n^{\nu}_0 \,dx_0^{\mu} \,d \Sigma_0= - p_0 dV_0.
\ee
Differentiating (\ref{2.4a}), (\ref{2.4b}) and applying (\ref{2.11}) one gets the first law of thermodynamics in the inertial frame $K$
\setcounter{orange}{1}
\renewcommand{\theequation} {\arabic{section}.\arabic{equation}\theorange}
\be
\label{2.14a}
dE= \gamma(v)dQ_0 +\gamma(v) \left( \delta L_0 + \frac{v^{\mu} v^{\rho}}{c^2} d{\cal T}_{0 \;\mu \rho}\right),
\ee
\addtocounter{orange}{1}
\addtocounter{equation}{-1}
\be
\label{2.14b}
dP^{\mu}= \gamma(v)\frac{v^{\mu}}{c^2}\delta Q_0 + \gamma(v)\frac{v^{\mu}}{c^2}\delta L_0 +
\frac{v^{\nu}}{c^2}\left[ \big(\gamma(v)-1 \big) \frac{v^{\mu}v^{\rho}}{v^2}-\eta^{\mu \rho}\right]d{\cal T}_{0 \;\nu \rho}.
\ee
\renewcommand{\theequation} {\arabic{section}.\arabic{equation}}
The notion `a reversible adiabatic process' should be independent of the choice of a system of frames. Consequently we  claim that: {\bf $\delta Q_0=0$ if and only if for any inertial frame $K$ the heat $\delta Q$ supplied to the thermodynamic system vanishes i.e. $\delta Q=0$}. Therefore (\ref{2.14a}) can be rewritten in the form 
\[
dE=\delta Q + \delta L
\]
\be
\label{2.15}
\delta Q = \gamma(v)\delta Q_0 \;\;, \;\; \delta L=  \gamma(v) \left( \delta L_0 + \frac{v^{\mu} v^{\rho}}{c^2} d{\cal T}_{0 \;\mu \rho}\right)
\ee
with $\delta Q$ standing for the heat supplied to the system and $\delta L$ denoting the thermodynamic work performed on the system with respect to the frame $K.$ Employing also (\ref{2.14b}) one can express the first law of relativistic thermodynamics in a $4$--D form
with $\delta Q$ standing for the heat supplied to the system and $\delta L$ denoting the thermodynamic work performed on the system with respect to the frame $K.$ Employing also (\ref{2.14b}) one can express the first law of relativistic thermodynamics in a $4$--form
\[
dP^j= \delta Q^j + \delta L^j\;\;, \;\; P^j=\left(P^1,P^2,P^3, \frac{E}{c}\right)= \left(\vec{P}, \frac{E}{c}\right),
\]
\be
\label{2.16}
\delta Q^j= \frac{\delta Q_0}{c}u^j\;\;,\;\; \delta L^{\mu}= \gamma(v)\frac{v^{\mu}}{c^2}\delta L_0 +
\frac{v^{\nu}}{c^2}\left[ \big(\gamma(v)-1 \big) \frac{v^{\mu}v^{\rho}}{v^2}-\eta^{\mu \rho}\right]d {\cal T}_{0 \;\nu \rho} \;\;,\;\; \delta L^4 = \frac{\delta L}{c}. 
\ee
Thus we arrive at the $4$--vector of heat $\delta Q^j$ and the $4$--object of relativistic thermodynamic work $\delta L^j.$ The work 
$\delta L^j$ is not a $4$--vector. 
The spatial part $\frac{\gamma(v)\delta Q_0}{c^2}\vec{v}$ of the $4$--vector of heat represents the change of momentum $d\vec{P}$ caused by the transfer of heat. The 4th component of $\delta Q^j$ is proportional (with the coefficient $\frac{1}{c}$) to  the change of energy following from the heat transfer.

Formulas (\ref{2.15}) or (\ref{2.16}) lead to the relativistic transformation of heat found by H. Ott \cite{hot} and independently by H.  Arzeli\`{e}s \cite{har} and C. M\o ller \cite{cmo}, \cite{cm2}
\be
\label{2.17}
\delta Q= \gamma(v) \delta Q_0
\ee
(see also \cite{cm3}, \cite{mr}).

This transformation rule differs dramatically from the one given by A. Einstein \cite{ae}, M. Planck \cite{max}, K. von Mosengeil \cite{kmo}, W . Pauli \cite{wp} and M. von Laue \cite{ml} 
\be
\label{2.18}
\delta Q^{(\rm Planck)}= \frac{1}{\gamma(v)} \delta Q_0.
\ee
(We must note that, as has been pointed out by Ch. Liu \cite{chl} after his studies of the letters between Einstein and Laue, A. Einstein in 1952 changed his opinion on the validity of (\ref{2.18}) in favour of (\ref{2.17})).

If we mention also that by some authors the heat supplied to be the system is considered to be a Lorentz invariant (see P. T. Landsberg and his collaborators \cite{ptl1}--\cite{ptl4} and N. G. van Kampen \cite{ngvk})
\be
\label{2.19}
\delta Q^{\rm (L)}= \delta Q_0
\ee
then one obtains  variety of approaches to the problem of relativistic transformation of heat. However, in our opinion according to (\ref{2.14a}) {\bf the most natural transformation rule} is given by Eq. (\ref{2.17}). Our conviction comes from the fact, that the infinitesimal heat $\delta Q$ and the infinitesimal work $\delta L$ must be of the form 
(\ref{2.15}).

The next question is: what about the relativistic transformation of temperature? Before we consider this question we write down the first law of relativistic thermodynamics (\ref{2.14a}), (\ref{2.14b}) in the case of ideal liquid. Inserting (\ref{2.8}) and (\ref{2.13}) into   (\ref{2.14a}) and (\ref{2.14b}) one gets
\setcounter{orange}{1}
\renewcommand{\theequation} {\arabic{section}.\arabic{equation}\theorange}
\be
\label{2.20a}
dE= \gamma(v) \delta Q_0 -pdV + \big(\gamma(v) \big)^2 \,\frac{v^2}{c^2}Vdp,
\ee
\addtocounter{orange}{1}
\addtocounter{equation}{-1}
\be
\label{2.20b}
d \vec{P}= \gamma(v) \frac{\vec{v}}{c^2}\delta Q_0 + \big( \gamma(v)\big)^2 \,\frac{\vec{v}}{c^2} V dp.
\ee
Hence
\renewcommand{\theequation} {\arabic{section}.\arabic{equation}}
\be
\label{2.21}
\delta L= -pdV + \big(\gamma(v) \big)^2 \,\frac{v^2}{c^2}Vdp\;\;, \;\;
\delta \vec{L}= \big( \gamma(v)\big)^2 \,\frac{\vec{v}}{c^2} V dp
\ee
(see \cite{cmo}, \cite{cm2}, \cite{cm3}, \cite{mr}).

Consider now the relativistic transformation of temperature. In nonrelativistic phenomenological thermodynamics we meet two main notions of temperature. The first one is the {\bf empirical temperature} introduced by the zeroth law of thermodynamics as a number determining the equivalence class of all thermodynamic systems being in thermal equilibrium. We assume that the zeroth law holds true also in relativistic thermodynamics. Therefore the {\bf empirical temperature in relativistic thermodynamics is a Lorentz invariant}. We will identify it with the  absolute temperature $T_0$ in the proper frame. The second notion, the {\bf absolute temperature}, is introduced by the second law of thermodynamics as the unique (up to a constant factor) integrating factor of the Pfaffian form $\delta Q_0$ depending only on the empirical temperature. This leads to the Clausius equality
\be
\label{2.22}
\oint \frac{\delta Q_0}{T_0}=0
\ee
for every cyclic reversible process. We assume that the second law of thermodynamics is in force in relativistic thermodynamics in any inertial frame. Consequently, the absolute temperature of a thermodynamic system moving with a constant velocity $\vec{v}$ with respect to the frame $K$ should be some smooth function of the form
\be
\label{2.23}
T=T(T_0,v)\;\;\;,\;\;\; \lim_{v \rightarrow 0} T(T_0,v)=T_0
\ee
and the Clausius equality in $K$ reads
\be
\label{2.24}
\oint \frac{\delta Q}{T}=0.
\ee
Substituting (\ref{2.17}) and (\ref{2.23}) into (\ref{2.24}) we obtain
\be
\label{2.25}
\oint \frac{\delta Q_0}{\big(\gamma(v)\big)^{-1}T(T_0,v)}=0.
\ee
Comparing (\ref{2.25}) with (\ref{2.22}) one arrives at the conclusion that 
\be
\label{2.26}
\big(\gamma(v)\big)^{-1}T(T_0,v)= b T_0
\ee
where $b$ is some constant. Taking the limit of Eq. (\ref{2.26}) for $v \rightarrow 0$ and applying (\ref{2.23}) we find that $b=1$ so finally
\be
\label{2.27}
T= \gamma(v)T_0.
\ee
This is the relativistic transformation of absolute temperature consistent with (\ref{2.17}) and the second law of relativistic thermodynamics for reversible processes. This transformation rule is dramatically different from the transformation law \cite{ae}--\cite{ml}
\be
\label{2.28}
T^{\rm (Planck)}= \big(\gamma(v)\big)^{-1} T_0
\ee 
following from (\ref{2.18}) or the transformation rule \cite{ptl1}--\cite{ptl4}
\be
\label{2.29}
T^{\rm (L)}=T_0
\ee  
consistent with (\ref{2.19}).

In the next section we are going to give some justification for (\ref{2.17}) and (\ref{2.27}).

Note also that by (\ref{2.12}), (\ref{2.17}) and (\ref{2.27}) one gets
\be
\label{2.30}
\delta Q = T dS_0
\ee
which means that according to Planck \cite{max} entropy is a Lorentz invariant 
\be
\label{2.31}
S=S_0
\ee
(see also \cite{cmo}, \cite{cm2}, \cite{rkp2}).


\section{From J\"{u}ttner distribution to phenomenological relativistic thermodynamics}

\label{section3}

\setcounter{equation}{0}

In 1911 F. J\"{u}ttner \cite{fj} proposed a relativistic counterpart of the famous Maxwell distribution for ideal gas in its proper inertial frame $K_0.$ The J\"{u}ttner distribution had been accepted for many years \cite{wp}, \cite{rkp1}--\cite{srg}, but from the 80's some authors have raised their objections to this distribution \cite{lph1}--\cite{jd}. 

The similar objections appear when the J\"{u}ttner distribution is generalized to the case when a vessel containing gas moves with a constant velocity $\vec{v}$ with respect to the laboratory frame $K.$ Hence, the natural question arises if the J\"{u}ttner formula and its generalization to the moving thermodynamic system (ideal gas) are correct. Nowadays we are not able to find any experimental solution of this question. Nevertheless, a partial answer comes from the recent outstanding works \cite{dc}--\cite{pea} where some computer simulations have been presented. In Refs. \cite{dc}, \cite{jd2} the computer simulation has been performed for a $1$--D ideal gas being a two-component  mixture of ideal gases. The numerical results show the perfect agreement with both: the $1$--D J\"{u}ttner distribution in the proper frame $K_0$ and its generalization to the moving vessel. In \cite{cr}--\cite{mh2} a $2$--D ideal gas is considered and the numerical results
of simulations are again in the perfect agreement with the $2$--D J\"{u}ttner distribution in the proper frame $K_0$. Moreover, in \cite{mh1}, \cite{mh2} the numerical results are shown to be in perfect agreement with the generalization of the J\"{u}ttner distribution for the moving vessel.

In \cite{pea} some theoretical considerations show the advantage of   the J\"{u}ttner distribution over other relativistic  distribution functions  and the $3$--D Monte Carlo simulations confirm the J\"{u}ttner distribution.

If so then it is an easy matter to get from these $1$--particle distributions the corresponding relativistic Gibbs distributions. Finally, the relativistic Gibbs distribution for an ideal gas contained in the vessel moving with a constant velocity $\vec{v}$  and $4$--velocity $u^j= \left(\gamma(v)\frac{\vec{v}}{c}, \gamma(v) \right)$ reads
\be
\label{3.1}
dw= \frac{1}{(2 \pi \hbar)^{3N}N!{\cal Z}} \exp \left\{ - \beta c u_j {\cal P}^j\right\} d^{3N}p \:d^{3N}q
\ee
where $N$ stands for the number of particles, $\beta= \frac{1}{kT_0} $ with $k$ being the Boltzmann constant, ${\cal P}^j= \left(\vec{\cal P}, \frac{\cal E}{c}\right)$ is the total momentum of the gas, $d^{3N}p \:d^{3N}q$ is the phase space volume element and ${\cal Z}$ denotes the partition function 
\be
\label{3.2}
{\cal Z}= \frac{V^N}{(2 \pi \hbar)^{3N}N!} \int_{{\mathbb R}^{3N}} \exp \left\{ - \beta c u_j {\cal P}^j\right\} d^{3N}p.
\ee
Remembering that the measure $d^{3}p$ transforms as follows \cite{rkp2}
\be
\label{3.3}
d^3 p= \gamma(v) \left( 1+ \frac{\vec{v}\cdot \vec{p}_0}{c \sqrt{\vec{p}^{\,2}_0+m^2c^2}}\right)d^3p_0
\ee
one gets
\[
{\cal Z}= \frac{V^N}{(2 \pi \hbar)^{3N}N!} \left(\int_{{\mathbb R}^{3}} \exp \left\{ - \beta c u_j p^{\,j}\right\} d^{3}p \right)^N=
\]
\[=
 \frac{V^N}{(2 \pi \hbar)^{3N}N!}  \left(\int_{{\mathbb R}^{3}} \exp \left\{ - \beta c \sqrt{\vec{p}^{\,2}_0+m^2c^2}
\right\} \left[\gamma(v) \left( 1+ \frac{\vec{v}\cdot \vec{p}_0}{c \sqrt{\vec{p}^{\,2}_0+m^2c^2}}\right)
 \right] d^3p_0
\right)^N=
\]
\be
\label{3.4}
=\frac{V_0^N}{(2 \pi \hbar)^{3N}N!} 
\left(\int_{{\mathbb R}^{3}} \exp \left\{ - \beta c \sqrt{\vec{p}^{\,2}_0+m^2c^2}
\right\}d^3p_0 \right)^N={\cal Z}_0
\ee
where, as in  Sec. 2, the subindex `0' corresponds to the proper frame $K_0$. Therefore,  the {\bf partition function ${\cal Z}$ is a Lorentz invariant} \cite{rkp2}. Then the entropy of the gas $S$ reads
\be
\label{3.5}
S= -k \Big< \ln \left( \frac{1}{\cal Z} \exp\left\{- \beta c u_j {\cal P}^j \right\}\right)\Big>= k \big(\ln {\cal Z} + \beta c u_j \big<{\cal P}^j \big> \big)
\ee 
with $\big< \cdot \big>$ denoting the expected value (average) with respect to  the Gibbs distribution (\ref{3.1}).

Analogous calculations as those done by R. K. Pathria \cite{rkp1}, \cite{rkp2} show that the gas pressure is a Lorentz invariant i.e. $p=p_0$ exactly as in the case of continuous medium from Sec. 2 and also that the following formulas, which are closely related to (\ref{2.9a}), (\ref{2.9b}), hold true
\setcounter{orange}{1}
\renewcommand{\theequation} {\arabic{section}.\arabic{equation}\theorange}
\be
\label{3.6a}
\big< {\cal E}\big>= \gamma(v)  \left(
\big< {\cal E}_0\big>_0+
 \frac{v^2}{c^2}p_0V_0 \right),
\ee
\addtocounter{orange}{1}
\addtocounter{equation}{-1}
\be
\label{3.6b}
\big< \vec{\cal P} \big>= 
\gamma(v)  \bigg(
\big< {\cal E}_0\big>_0+
 p_0V_0 \bigg) \frac{\vec{v}}{c}
\ee
\renewcommand{\theequation} {\arabic{section}.\arabic{equation}}
(see also \cite{chn}).
Hence, as before, $\left(\big< \vec{\cal P} \big>, \frac{\big< {\cal E}\big>+pV}{c} \right)$ i.e. the average total momentum $\big< \vec{\cal P} \big>$ and the enthalpy of the gas divided by $c$ namely  $\frac{\big< {\cal E}\big>+pV}{c}$ constitute a $4$--vector. 

Employing (\ref{3.6a}) and (\ref{3.6b}) one quickly finds 
\be
\label{3.7}
u_j \big< \vec{\cal P}^j \big>= \frac{\big< {\cal E}_0\big>_0}{c}.
\ee
Thus, although $\big< \vec{\cal P}^j \big>$ is not a $4$--vector, the `scalar product' $u_j \big< \vec{\cal P}^j \big>$ is a Lorentz invariant.

Inserting (\ref{3.4}) and (\ref{3.7}) into (\ref{3.5}) we obtain the result
\be
\label{3.8}
S= k \big( \ln {\cal Z}_0 + \beta  \big< {\cal E}_0\big>_0 \big)=S_0
\ee
which confirms the Planck formula (\ref{2.31}).

Differentiating (\ref{3.6a}) and (\ref{3.6b}) and performing the same calculations as in  Sec. 2 one arrives at the first law of relativistic thermodynamics for ideal gas in the form (\ref{2.20a}) and (\ref{2.20b}) with
\be
\label{3.9}
E:= \big< {\cal E} \big>\;\;\;, \;\;\; \vec{\cal P}:= \big< \vec{\cal P} \big>.
\ee
Hence, again we are led to the relativistic transformations of heat (\ref{2.17}) and absolute temperature (\ref{2.27}).

However, we are going to give a slightly deeper insight into the problem by applying some statistical considerations which are well known in nonrelativistic  statistical physics and which lead from the Gibbs distributions to the first law of thermodynamics \cite{ldl}.

To this end we rewrite the relativistic Gibbs distribution (\ref{3.1}) in a quantum form
\be
\label{3.10}
w_n= \frac{1}{\cal Z} \exp \big\{ - \beta c u_j {\cal P}^j_n \big\}\;\;\;, \;\;\; n=1,2,\ldots
\ee
where the subindex `$\,n\,$' denotes a quantum state. One assumes that the total $4$--momentum eigenvalues ${\cal P}^j_n$ depend on some external thermodynamic parameters $\lambda_1, \ldots, \lambda_s.$ Differentiating the formula
\be
\label{3.11}
\sum_{n=1}^{\infty}w_n=1
\ee
with respect to $T_0,\lambda_1, \ldots, \lambda_s, $ and performing some simple manipulations we get
\be
\label{3.12}
u_j d\big< {\cal P}^j\big>= \frac{1}{k \beta c}dS + u_j \big< d{\cal P}^j\big>,
\ee
\[
\big< {\cal P}^j\big>:= \sum_{n=1}^{\infty}w_n {\cal P}^j_n \;\;\;, \;\;\;  \big< d{\cal P}^j\big>
:= \sum_{n=1}^{\infty}w_n \left( \sum_{l=1}^s \frac{\partial {\cal P}^j_n}{\partial \lambda_l} d \lambda_l\right).
\]
Keeping in mind that $u_j=\Big(-\,\gamma(v)\, \frac{\vec{v}}{c}, \gamma(v) \Big)$ and ${\cal P}^j_m=\left(\vec{\cal P}_m,\frac{{\cal E}_m}{c}\right)$ one can rewrite (\ref{3.12}) in the following form
\be
d \big< {\cal E}\big>= \frac{T_0}{\gamma(v)}dS + \vec{v}\cdot \big(d \big< \vec{\cal P}\big> - \big< d\vec{\cal P}\big>\big)+
\big< d{\cal E}\big>.
\label{3.13}
\ee
Assuming that, analogously as in the proper frame $K_0$ the statistical definition of thermodynamic work $\delta L$ in any inertial frame reads
\be
\label{3.14}
\delta L= \big< d{\cal E}\big>
\ee 
(see \cite{sch}) and comparing (\ref{2.15}) with (\ref{3.13}) we obtain
\be
\label{3.15}
\delta Q= \frac{T_0}{\gamma(v)}dS + \vec{v}\cdot \big(d \big< \vec{\cal P}\big> - \big< d\vec{\cal P}\big>\big).
\ee
From (\ref{3.15}) with (\ref{2.12}) and (\ref{3.8}) it follows that statistical considerations give the transformation rule 
(\ref{2.18}) for heat if either $\vec{v}\cdot \big(d \big< \vec{\cal P}\big> - \big< d\vec{\cal P}\big>\big)=0$ or 
one decides to change Def. (\ref{3.14}) of $\delta L$ into
\be
\label{3.16}
\delta L \rightarrow \delta L'= \big< d{\cal E}\big> + \vec{v}\cdot \big(d \big< \vec{\cal P}\big> - \big< d\vec{\cal P}\big>\big).
\ee
The latter possibility explains from the statistical physics point of view the reason why A. Einstein in 1907, M. Planck in 1908 and others found  transformation formula (\ref{2.18}). Employing the detailed phenomenological analysis of C. M\o ller \cite{cmo} one can expect that
\be
\label{3.17}
\big< d\vec{\cal P}\big>= \big( \gamma(v)\big)^2 \frac{\vec{v}}{c^2} V dp.
\ee 
(As yet we have not been able to derive (\ref{3.17}) using laws of statistical physics only).

Then from (\ref{2.20b}) with (\ref{3.9}) and (\ref{3.17}) we have
\be
\label{3.18}
\vec{v}\cdot \big(d \big< \vec{\cal P}\big> - \big< d\vec{\cal P}\big>\big)=\gamma(v) \frac{v^2}{c^2} \delta Q_0.
\ee

Finally, inserting (\ref{3.18}) into (\ref{3.15}) one arrives at the transformation rule (\ref{2.17}) of H. Ott, H. Arzeli\`{e}s and C. M\o ller.

Moreover, (\ref{2.20a}) and (\ref{2.20b}) with (\ref{3.9}), (\ref{2.12}), (\ref{2.16}), (\ref{3.13})--(\ref{3.18}) lead to the Lorentz invariant formulation of the first law of relativistic thermodynamics
\be
\label{3.19}
d \big< {\cal P}^j\big> - \big< d{\cal P}^j\big>= T^j dS\;\;\;, \;\;\;T^j:=\frac{T_0}{c}u^j.
\ee
$T^j$ is the $4$--vector of temperature \cite{cmo}, \cite{cm2}. One quickly finds that by (\ref{2.27}) 
\be
\label{3.20}
T^4=T_4= \gamma(v)\frac{T_0}{c}= \frac{T}{c}.
\ee
(Analogous approach to the relation between relativistic statistical physics and relativistic phenomenological thermodynamics had been presented by P. G. Bergmann in 1951 \cite{pgb} (see also \cite{md}) many years before  computer simulations were done by D. Cubero {it et al.} \cite{dc}, \cite{jd2} and C. Rasinariu \cite{cr}).

It seems that the statistical considerations of this section, which follow directly from the relativistic Gibbs distribution (\ref{3.1}) and consequently,  from the J\"{u}ttner distribution, give evidence that the correct transformation rules for heat and temperature are given by (\ref{2.17}) and (\ref{2.27}) respectively.
\section{On some identities in relativistic thermodynamics. Operational definition of the absolute temperature $T$}

\label{section4}

\setcounter{equation}{0}

First, using (\ref{2.12}),  (\ref{2.27}) and (\ref{2.31}) we rewrite (\ref{2.20a}) and (\ref{2.20b}) in the form
\setcounter{orange}{1}
\renewcommand{\theequation} {\arabic{section}.\arabic{equation}\theorange}
\be
\label{4.1a}
dE= TdS -pdV + \big(\gamma(v) \big)^2 \,\frac{v^2}{c^2}Vdp,
\ee
\addtocounter{orange}{1}
\addtocounter{equation}{-1}
\be
\label{4.1b}
d \vec{P}= \frac{\vec{v}}{c^2} T dS + \big( \gamma(v)\big)^2 \,\frac{\vec{v}}{c^2} V dp.
\ee
\renewcommand{\theequation} {\arabic{section}.\arabic{equation}}
From (\ref{4.1a}) one quickly finds
\be
\label{4.2}
d \Big(E - \big( \gamma(v)\big)^2 \,\frac{v^2}{c^2} V p \Big)= T dS - \big( \gamma(v)\big)^2 \, p dV. 
\ee
Hence
\setcounter{orange}{1}
\renewcommand{\theequation} {\arabic{section}.\arabic{equation}\theorange}
\be
\label{4.3a}
T = \left( \frac{\partial E}{\partial S}\right)_V - \big( \gamma(v)\big)^2 \,\frac{v^2}{c^2} V \left(\frac{\partial p}{\partial S} \right)_V,
\ee
\addtocounter{orange}{1}
\addtocounter{equation}{-1}
\be
\label{4.3b}
p - \big( \gamma(v)\big)^2 \,\frac{v^2}{c^2} V \left(\frac{\partial p}{\partial V} \right)_S = - \left( \frac{\partial E}{\partial V}\right)_S
\ee
\renewcommand{\theequation} {\arabic{section}.\arabic{equation}}
and
\be
\label{4.4}
\left( \frac{\partial T}{\partial V}\right)_S= - \big( \gamma(v)\big)^2 \left( \frac{\partial p}{\partial S}\right)_V.
\ee
Moreover, from (\ref{4.2}) and (\ref{4.1b}) we obtain
\be
\label{e4.4}
T= \big( \gamma(v)\big)^2 \left(\frac{\partial E}{\partial S} \right)_{V,\vec{P}} 
\ee
(compare with formula (9) from the paper \cite{rkp2}).

As we showed in Sec. 2, the sum $E+pV$ transforms according to the rule (\ref{2.10}). We identify it with enthalpy $H.$ Hence we see that
\be
\label{e1,4.4}
H= \gamma(v)H_0.
\ee 
Applying (\ref{4.2}) and definition of enthalpy we get
\be
\label{e2.4.4}
T= \left( \frac{\partial H}{\partial S}\right)_{p}.
\ee

Define the free energy $F$
\be
\label{4.5}
F:= E-TS- \big( \gamma(v)\big)^2 \,\frac{v^2}{c^2}p V= \gamma(v)F_0
\ee
(the last equality follows from (\ref{2.27}), (\ref{2.31}) and (\ref{2.9a})). $F_0= E_0 - T_0 S_0$ is the free energy of the system in its proper frame $K_0.$ Inserting (\ref{4.5}) into (\ref{4.2}) we get
\be
\label{4.6}
dF=-SdT - \big( \gamma(v)\big)^2 pdV. 
\ee
From (\ref{4.6}) one finds
\be
\label{4.7}
S=- \left( \frac{\partial F}{\partial T}\right)_V \;\;, \;\;  \big( \gamma(v)\big)^2 p=-\left( \frac{\partial F}{\partial V}\right)_T\;\; ,\;\; E=F- T \left( \frac{\partial F}{\partial T}\right)_V -\frac{v^2}{c^2} V \left( \frac{\partial F}{\partial V}\right)_T
\ee
and one of the relativistic Maxwell identities
\be
\label{4.8}
\left( \frac{\partial S}{\partial V}\right)_T= \big( \gamma(v)\big)^2 \left( \frac{\partial p}{\partial T}\right)_V.  
\ee
We define the Gibbs function $G$ in a standard way
\be
\label{4.9}
G:= E-TS+pV= \gamma(v) G_0
\ee
(use (\ref{2.27}), (\ref{2.31}) and (\ref{2.9a})), with $G_0=E_0 - T_0 S_0 + p_0 V_0.$

Differentiating (\ref{4.9}) and substituting (\ref{4.1a}) we obtain the formula
\be
\label{4.10}
dG= -SdT + \big( \gamma(v)\big)^2 Vdp
\ee
from which we get the relations
\be
\label{4.11}
S= - \left( \frac{\partial G}{\partial T}\right)_p \;\; , \;\; 
\big( \gamma(v)\big)^2 V =\left( \frac{\partial G}{\partial p}\right)_T \;\;,\;\;
E= G - T \left( \frac{\partial G}{\partial T}\right)_p - \big( \gamma(v)\big)^{-2} p \left( \frac{\partial G}{\partial p}\right)_T
\ee
and the Maxwell identity
\be
\label{4.12}
\left( \frac{\partial S}{\partial p}\right)_T= - \big( \gamma(v)\big)^2 \left( \frac{\partial V}{\partial T}\right)_p.
\ee
We conclude that the
thermodynamic potentials $H,F,G$ are defined like in a rest frame and satisfy the same transformation rule.

In the celebrated monograph of L. D. Landau and E. M. Lifshitz \cite{ldl} the formula (\ref{4.12}) with $\gamma(v)=1$ is employed to find the absolute temperature as a function of empirical temperature in nonrelativistic thermodynamics. Modifying slightly those considerations we can obtain an analogous result in relativistic thermodynamics.

Namely, writing $\left( \frac{\partial V}{\partial T}\right)_p= \left( \frac{\partial V}{\partial T_0}\right)_p \frac{dT_0}{dT}$ and remembering that $p=p_0, S=S_0,$ from (\ref{4.12}) one gets
\be
\label{4.13}
\frac{dT_0}{dT}= - \big( \gamma(v)\big)^2 \frac{T_0\left( \frac{\partial V}{\partial T_0}\right)_p}{T_0\left( \frac{\partial S_0}{\partial p_0}\right)_{T_0}}.
\ee
This formula gives the derivative $\frac{dT_0}{dT}$ in terms of measurable quantities. Indeed, 
$\left( \frac{\partial V}{\partial T_0}\right)_p$ is determined by the change of the volume $V$ under the change of $T_0$ for the constant pressure $p.$ Moreover, $T_0 \left( \frac{\partial S_0}{\partial p_0}\right)_{T_0}$ is determined by the heat $\delta Q_0$ supplied to the thermodynamic system so that under the change of pressure $p_0$ the temperature $T_0$ remains unchanged 
\[
T_0 \left( \frac{\partial S_0}{\partial p_0}\right)_{T_0}= \lim_{\Delta p_0 \rightarrow 0} \left( \frac{Q_0}{\Delta p_0}\right)_{T_0}.
\]
Assuming also that 
\be
\label{4.14}
\lim_{T_0 \rightarrow 0} T=0
\ee
what is consistent with the third law of thermodynamics (note (\ref{2.31})) one finds that formula (\ref{4.13}) determines $T=T(T_0,v).$

Therefore, formula (\ref{4.13}) indicates an operational definition of the absolute temperature $T.$

Another operational definition of the absolute temperature $T$ is provided by the Clausius equation (\ref{2.24}) when applied to an appropriate thermodynamic engine. This idea was presented by  C. M\o ller \cite{cm2} and then considered in Refs. \cite{mp}, \cite{mr}. We remind briefly main points of  M\o ller's construction. Let $R$ and $R_0$ be two reservoirs. $R_0$ is at rest with respect to the inertial frame $K_0$ and $R$ is at rest with respect to the system $K.$ The absolute temperature of $R_0$ with respect to $K_0$ is $T_0$ and it is equal to the absolute temperature of the reservoir $R$ with respect to the system $K.$ 

Let us consider    a thermodynamic engine operating between the reservoirs $R_0 $ and $R$ according to some Carnot cycle. The whole process is analyzed from the frame $K_0.$ First, the engine absorbs isothermally the amount of heat $Q_0$ from the reservoir $R_0$ at the absolute temperature $T_0,$ being at rest with respect to the reference system $K_0.$ Then the engine is accelerated adiabatically to the velocity  of the frame $K$ equal to $- \vec{v}.$ The amount of heat $Q$ (with respect to the system $K_0$ i.e. $Q_0$ with respect to the frame $K$) is released isothermally from the engine to the reservoir $R$ at the temperature $T$ (with respect to the reference frame $K_0$ i.e. the temperature $T_0$ with respect to the system $K$). The final step is adiabatic deceleration of the engine so that it returns to its initial state.

The efficiency $\eta$ of this cycle reads
\be
\label{4.15}
\eta= 1 - \frac{Q}{Q_0}= 1 - \frac{T}{T_0}.
\ee
Consequently, by measuring $\eta$ one finds  transformation rules of heat and of temperature. Thus {\bf both (\ref{2.17}) and (\ref{2.27}) can be verified experimentally}. (About the relation between the absolute temperature and the efficiency of the Carnot cycle see also \cite{kh}).  

Taking the scalar product of both sides of Eq. (\ref{4.1b}) with $\vec{v}$ and subtracting the result from (\ref{4.1a}) we get 
\be
\label{4.16}
dE= \big( \gamma(v)\big)^{-1}T_0 dS -pdV + \vec{v} \cdot d \vec{P}.
\ee  
In the case of an ideal gas one can employ (\ref{3.18}) and then (\ref{4.16}) is brought to the form
\be
\label{4.17}
dE=  \gamma(v)T_0 dS -pdV + \vec{v} \cdot \big< d \vec{P} \big>.
\ee
The quantity $\big< d \vec{P} \big>$ is interpreted as the {\bf mechanical momentum } supplied to the system \cite{cmo}, \cite{cm2}. With this interpretation the formula (\ref{4.17}) is true also for an ideal liquid. ($\big< d \vec{P} \big>$ corresponds to $\Delta \vec{J}$ in Refs. \cite{cmo}, \cite{cm2}). Observe that in general $\big< d \vec{P} \big>$ is not the differential of any thermodynamic function. 

Returning to identities in relativistic thermodynamics we should note that we have considered them under the assumption that the number of particles $N$ and the velocity of the thermodynamic system $\vec{v} $ are constant. Consequently, many of our formulas differ from the respective relations of relativistic thermodynamics presented in Ref. \cite{caf} (compare, for example, our (\ref{4.3a}) with Eq.(15) of \cite{caf} ).

It is a straightforward matter to generalize all results to the case when $N$ and $\vec{v}$ are not constant. Differentiating (\ref{2.9a}) and (\ref{2.9b}) and employing the first law of thermodynamics in the rest frame
\be
\label{4.15.1}
dE_0=T_0 dS_0-p_0 dV_0 + \mu_0 dN_0
\ee
where $\mu_0$ is the chemical potential with respect to the rest frame and $N_0=N,$ after some simple manipulations one gets
\setcounter{orange}{1}
\renewcommand{\theequation} {\arabic{section}.\arabic{equation}\theorange}
\be
\label{e4.16a}
dE= TdS -pdV + \big(\gamma(v) \big)^2 \,\frac{v^2}{c^2}Vdp+ \big(\gamma(v) \big)^3 \frac{E_0+p_0 V_0}{c^2} \vec{v} \cdot d \vec{v} + \mu dN ,
\ee
\addtocounter{orange}{1}
\addtocounter{equation}{-1}
\be
\label{e4.16b}
d \vec{P}= \frac{\vec{v}}{c^2} T dS + \big( \gamma(v)\big)^2 \,\frac{\vec{v}}{c^2} V dp + \big( \gamma(v)\big)^3 \,
\frac{E_0+p_0 V_0}{c^2}\left[ d \vec{v}+\frac{1}{c^2} \vec{v} \times (\vec{v} \times d \vec{v})\right]+\mu \frac{\vec{v}}{c^2} dN,
\ee
\renewcommand{\theequation} {\arabic{section}.\arabic{equation}}
where $T$ is given by (\ref{2.27}) and
\be
\label{e4.17}
\mu= \gamma(v)\mu_0.
\ee
From (\ref{e4.16a}) and (\ref{e4.16b}) one infers that analogously to the $4$--vector of temperature $T^j$ (see (\ref{3.19}) and (\ref{3.20})) we can introduce the $4$--vector of chemical potential
\be
\label{e4.18}
\mu^j= \frac{\mu_0}{c}u^j \Rightarrow u^4= \gamma(v) \frac{\mu_0}{c}= \frac{\mu}{c}.
\ee
Then, from (\ref{e4.16a}) and (\ref{e4.16b}) we quickly find that the thermodynamic work $\delta L$ reads
\be
\label{e4.19}
\delta L= -pdV + \big(\gamma(v) \big)^2 \,\frac{v^2}{c^2}Vdp+ \big(\gamma(v) \big)^3 \frac{E_0+p_0 V_0}{c^2} \vec{v} \cdot d \vec{v}
\ee
and $\delta \vec{L}$ (compare with (\ref{2.21})) is 
\be
\label{e4.20}
\delta \vec{L}= \big( \gamma(v)\big)^2 \,\frac{\vec{v}}{c^2} V dp + \big( \gamma(v)\big)^3 \,
\frac{E_0+p_0 V_0}{c^2}\left[ d \vec{v}+\frac{1}{c^2} \vec{v} \times (\vec{v} \times d \vec{v})\right].
\ee
It is an easy matter to generalize all results obtained before. For example, in the present case the identity (\ref{4.10}) reads
\be
\label{e4.21}
dG= - SdT + \big( \gamma(v)\big)^2 \,Vdp + \big(\gamma(v) \big)^3 \frac{E_0+p_0 V_0}{c^2} \vec{v} \cdot d \vec{v} + \mu dN.
\ee
Hence
\be
\label{e4.22}
\mu= \left( \frac{\partial G}{\partial N}\right)_{T,p,\vec{v}}.
\ee
Writing $G$ in the form
\be
\label{e4.23}
G= N g (T,p,\vec{v}) \Rightarrow g (T,p,\vec{v})= \left( \frac{\partial G}{\partial N}\right)_{T,p,\vec{v}}
\ee
and comparing (\ref{e4.22}) with (\ref{e4.23}) we get the formula well known in nonrelativistic thermodynamics
\be
\label{e4.24}
G=N \mu.
\ee
Then the potential $\Omega :=F- \mu N$ takes the form (use (\ref{4.5}) and (\ref{4.9}))
\be
\label{e4.25}
\Omega= F - \mu N= F-G= E-TS- \big( \gamma(v)\big)^2 \,\frac{v^2}{c^2}p V - \big( E-TS+pV \big)=- \big( \gamma(v)\big)^2 \,p V.
\ee
Of course
\be
\label{e4.26}
\Omega= -\gamma(v) p_0 V_0= \gamma(v) \Omega_0
\ee
and
\be
\label{e4.27}
N= - \left( \frac{\partial \Omega}{\partial \mu}\right)_{T,V,\vec{v}}=
- \left( \frac{\partial \Omega_0}{\partial \mu_0}\right)_{T_0,V_0}=N_0.
\ee
Now we can easily find an identity which has been used by many authors as a starting point for relativistic thermodynamics \cite{cmo}, \cite{rkp2}, \cite{caf}. Multiplying (\ref{e4.16b}) by $\vec{v}$ and subtracting from (\ref{e4.16a}) one gets
\be
\label{e4.28}
dE= \big( \gamma(v)\big)^{-2}TdS-pdV+ \vec{v} \cdot d \vec{P} + \big( \gamma(v)\big)^{-2} \mu dN.
\ee
In this approach the energy $E$ is considered as a function of $(S,V, \vec{P},N).$ 
A motion of the system with velocity $\vec{v}$ increases its number of degrees of freedom. New conjugated parameters $\vec{v}$ and $\vec{P}$ appear. Variables $S,V,\vec{P}$ and $N$ are natural variables for energy $E.$ Unfortunately, neither $\left(\frac{\partial E}{\partial S} \right)_{V, \vec{P},N}$ is the temperature nor  
$\left(\frac{\partial E}{\partial N} \right)_{S,V, \vec{P}}$ is the chemical potential.

Then $F=F(T,V,\vec{P},N)$ and 
$\Omega= \Omega(T,V, \vec{P},\mu)$. Note that from (\ref{2.9b}) and (\ref{2.10}) we have
\setcounter{orange}{1}
\renewcommand{\theequation} {\arabic{section}.\arabic{equation}\theorange}
\be
\label{e4.29a}
\gamma(v)= \frac{\sqrt{(E_0+p_0 V_0)^2+c^2 \vec{P}^2}}{E_0 +p_0 V_0},
\ee
\addtocounter{orange}{1}
\addtocounter{equation}{-1}
\be
\label{e4.29b}
E+pV= \sqrt{(E_0+p_0 V_0)^2+c^2 \vec{P}^2},
\ee
\addtocounter{orange}{1}
\addtocounter{equation}{-1}
\be
\label{e4.29c}
\vec{v}= \frac{c^2 \vec{P}}{\sqrt{(E_0+p_0 V_0)^2+c^2 \vec{P}^2}}.
\ee
\renewcommand{\theequation} {\arabic{section}.\arabic{equation}}
Keeping in mind that $S=S_0, p=p_0(S_0,V_0,N_0), V= \big( \gamma(v)\big)^{-1} V_0, N=N_0$ and $E_0=E_0(S_0,V_0,N_0),$ from (\ref{e4.29a}) and (\ref{e4.29c}) one finds that $\vec{v}$ as a function of $(S,V, \vec{P},N).$  This enables us to pass from (\ref{e4.16a}) where $E=E(S,V, \vec{v},N)$ to (\ref{e4.28}) where $E=E(S,V, \vec{P},N).$  

In distinguished papers on relativistic thermodynamics by R.K. Pathria \cite{rkp2} and H. Callen and G. Horwitz  \cite{chn} it has been shown that it is more convenient and natural to use the independent variables $(S,p,N)$ rather than $(S,V,N)$ and the enthalpy $H=E+pV$ instead of energy $E.$ This is because $S,p$ and $N$ are Lorentz invariants (but $V$ is not) and $(\vec{P}, \frac{H}{c})$ constitute a $4$--vector (see Sec. 2).

Adding $d(pV)$ to both sides of (\ref{e4.16a})  one obtains
\be
\label{e4.30}
dH= TdS + \big(\gamma(v) \big)^2 V dp + \big(\gamma(v) \big)^3 \frac{H_0}{c^2}\vec{v} \cdot d \vec{v} + \mu dN.
\ee  
Here $H= H(S,p,\vec{v},N).$

Analogously, adding $d(pV)$ to both sides of (\ref{e4.28}) we have
\be
\label{e4.31}
dH= \big( \gamma(v)\big)^{-2}TdS + Vdp + \vec{v} \cdot d \vec{P} + \big( \gamma(v)\big)^{-2} \mu dN,
\ee
where $H= H(S,p,\vec{P},N)$ is given by (\ref{e4.29b})
\be
\label{e4.32}
H= \sqrt{(E_0+p_0 V_0)^2+c^2 \vec{P}^2}= \sqrt{H_0^2+c^2 \vec{P}^2}.
\ee
In particular, from (\ref{e4.31}) with (\ref{e4.32}) one finds
\setcounter{orange}{1}
\renewcommand{\theequation} {\arabic{section}.\arabic{equation}\theorange}
\be
\label{e4.33a}
V= \left( \frac{\partial H}{\partial p}\right)_{S, \vec{P},N}= V_0 \frac{H_0}{H}= V_0 \frac{E_0 +p_0 V_0}{\sqrt{(E_0+p_0 V_0)^2+c^2 \vec{P}^2}},
\ee
\addtocounter{orange}{1}
\addtocounter{equation}{-1}
\be
\label{e4.33b}
\vec{v}= \left( \frac{\partial H}{\partial \vec{P}}\right)_{S, p,N}=  \frac{c^2 \vec{P}}{H}= \frac{c^2 \vec{P}}{\sqrt{(E_0+p_0 V_0)^2+c^2 \vec{P}^2}}.
\ee
\renewcommand{\theequation} {\arabic{section}.\arabic{equation}}
As it can be seen, (\ref{e4.33b}) is exactly (\ref{e4.29c}). 

We end the present section with an important remark. Thermodynamic identity (\ref{e4.16a}) with (\ref{e4.19}) lead to the Ott--Arzeli\`{e}s--C. M\o ller transformation of absolute temperature (\ref{2.27}) and to the transformation of chemical potential as given by (\ref{4.17}). However, the identity (\ref{e4.28}) suggests the Planck transformation of temperature (\ref{2.28}) $T^{\rm (Planck)}= \big(\gamma(v)\big)^{-2} T= \big(\gamma(v)\big)^{-1} T_0$ and the transformation of the chemical potential given by
$\mu^{\rm (Planck)}=  \big(\gamma(v)\big)^{-2} \mu= \big(\gamma(v)\big)^{-1} \mu_0 $ under the assumption that the thermodynamic work reads
\be
\label{e4.35}
\delta L'= -pdV + \vec{v} \cdot d \vec{P}.
\ee
Substituting  $\vec{v} \cdot d \vec{P}$ calculated from  (\ref{e4.16b}) we obtain
\be
\label{e4.36}
\delta L'= -pdV + \big(\gamma(v) \big)^2 \frac{v^2}{c^2}V dp+
\big(\gamma(v) \big)^3 \frac{E_0+p_0 V_0}{c^2}\vec{v} \cdot d \vec{v} + \frac{v^2}{c^2}TdS + \frac{v^2}{c^2} \mu dN.
\ee
Comparing (\ref{e4.36}) with (\ref{e4.19}) one concludes that the difference between the transformation rules of the absolute temperature and the chemical potential derived from identity (\ref{e4.16a}) and the rules derived from (\ref{e4.28}) is caused by the fact that $\delta L'$ given by (\ref{e4.35}) cannot be interpreted as the relativistic thermodynamic work. In fact 
$\delta L'$ consists of three groups of terms.
The  term $-pdV + \big(\gamma(v) \big)^2 \frac{v^2}{c^2}V dp+ \big(\gamma(v) \big)^3 \frac{E_0+p_0 V_0}{c^2}\vec{v} \cdot d \vec{v}$ defines the thermodynamic work $\delta L,$ the component equal to $\frac{v^2}{c^2}TdS$ corresponds to a part of relativistic heat supplied to the system and, finally, the  part $\frac{v^2}{c^2} \mu dN$ is an element of $dE$ which corresponds to the increment $dN$ of the number of particles.

If $dN=0$ then a statistical interpretation  of $\delta L'$ is given by (\ref{3.16}) with (\ref{3.14}).

\section{Statistical thermometers}

\label{section5}
\setcounter{equation}{0}

In the preceding sections we have argued that in relativistic phenomenological thermodynamics one meets two operationally well defined notions of temperature: the empirical temperature which can be identified with the proper absolute temperature $T_0$ and which is a Lorentz invariant, and the absolute temperature $T$ which follows from the second law of thermodynamics for reversible processes. According to the Clausius equation (\ref{2.24}) this absolute temperature can be experimentally determined as indicated by  measurement of  the efficiency $\eta$ (\ref{4.15}) of an appropriate thermodynamic engine. We have also argued that the transformation rule for $T$ is given by  formula (\ref{2.27}) and, consequently, $T$ can be experimentally determined as it is indicated by Eq. (\ref{4.13}).

Now we are going to consider the question, how one can define temperature in relativistic statistical thermodynamics of an ideal gas. To this end we employ the results of Sections \ref{section2}, \ref{section3} and \ref{section4} to find some thermodynamic functions for the ideal gas. 

First, performing integration in (\ref{3.4}) one gets
\be
\label{5.1}
{\cal Z}= {\cal Z}_0= \frac{1}{N!} \left( \frac{m^2c V_0}{2 \pi^2 \hbar^3 \beta} K_2(\beta m c^2)\right)^N, 
\ee
where $K_2(x)$ denotes the modified Bessel function of the second kind of order 2. Its integral representation is of the form 
\be
\label{5.2}
K_{\nu}(x)= \int_{0}^{\infty} e^{-x \cosh t} \cosh \nu t \,dt, \;\;\; x >0.
\ee
Inserting (\ref{5.1}) into the thermodynamic equation
\be
\label{5.3}
\big<{\cal E}_0 \big>_0= - \left( \frac{\partial \ln {\cal Z}_0}{\partial \beta}\right)_{V_0}
\ee
and employing the recurrence relation
\be
\label{5.4}
\frac{d K_{\nu}(x)}{dx} + \frac{\nu}{x}= - K_{\nu-1}(x)
\ee
we easily obtain
\be
\label{5.5}
E_0 \equiv \big<{\cal E}_0 \big>_0= \frac{3N}{\beta} + N m c^2 \frac{K_1(\beta m c^2)}{K_2(\beta m c^2)}.
\ee
Substituting (\ref{5.1}) and (\ref{5.5}) into (\ref{3.8}) and for large $N$ applying the well known Stirling formula  
$\ln N! \approx N \ln \frac{N}{e}$ one gets
\be
\label{5.6}
S=S_0 = kN \left[ \ln \left(\frac{m^2 c}{2 \pi^2 \hbar^3 \beta} \frac{V_0}{N} K_2(\beta m c^2) \right) + \beta m c^2
\frac{K_1(\beta m c^2)}{K_2(\beta m c^2)} +4
\right].
\ee
From (\ref{3.6a}), (\ref{3.8}), (\ref{3.9}) and (\ref{4.5}) we have
\be
\label{5.7}
F= - kT \ln {\cal Z}= \gamma(v) \big(- kT_0 \ln {\cal Z}_0 \big)= \gamma(v) F_0.
\ee
Then from (\ref{4.7}) with (\ref{5.1}) one obtains
\be
\label{5.8}
p=p_0= k T_0 \left( \frac{\partial \ln {\cal Z}_0}{\partial V_0} \right)_{T_0}= \frac{NkT_0}{V_0} \Longrightarrow
p_0 V_0 = Nk T_0.
\ee
Note that the formula (\ref{5.7}) justifies the name `free energy' for $F$ (see Section \ref{section4}, Eq. (\ref{4.5})).

We are ready to propose a model of a `natural' statistical thermometer. 
We start from the observation that the average energy of the ideal gas in its proper frame $K_0$ in nonrelativistic thermodynamics is given by
\be
\label{5.9}
E_0^{\rm nonrel.}= \lim_{c \rightarrow \infty} \big( \big< {\cal E}_0\big>_0 - Nmc^2\big)
\ee
where $\big< {\cal E}_0\big>_0 $ is defined by (\ref{5.5}). Applying the principle of equipartition of energy
\be
\label{5.9a}
\left< p^{\,\mu} \frac{\partial H^{\rm nonrel.}}{\partial p^{\,\mu}}\right>= kT_0,\;\;\; {\rm (\, do \;\; not \;\; sum \;\; over \;\;} \mu \, {\rm !})
\ee
where $H^{\rm nonrel.}$ denotes the Hamilton function of a system, we immediately obtain 
\be
\label{5.11}
T_0 = \frac{E_0^{\rm nonrel.}}{\frac{3}{2}Nk}. 
\ee
By analogy to (\ref{5.11}) one can introduce the {\bf statistical temperature} in relativistic thermodynamics.

The  generalized principle of equipartition of energy in a moving frame  for a fixed index $j$ is of the form (see also \cite{ptl2}, \cite{mh1})
\be
\label{5.11a}
\Big< p^{\,\mu} \frac{\partial H}{\partial p^{\,\mu}}\Big>= \frac{kT_0}{\gamma}  + v^{\,\mu} \big< p^{\,\mu} \big> \;\;\; {\rm (\, do \;\; not \;\; sum \;\; over \;\;} \mu \, {\rm !}).
\ee
Then in a similar way  to (\ref{5.11}) we propose
\be
\label{5.11b}
T^{\rm(st)}:= \frac{1}{3Nk}\left(\big< {\cal E} \big> - Nmc^2 \left< \frac{mc^2}{H_{\rm particle}}\right> - \vec{v} \cdot \big< \vec{\cal P}\big> \right) .
\ee
Thus $T^{\rm(st)}=T^{\rm (Planck)}=\frac{T_0}{\gamma(v)}.$  The statistical temperature $T^{\rm(st)}$ measures  the average kinetic energy modulo the contribution of translatory motion of the vessel. The term $\left< \frac{mc^2}{H_{\rm particle}}\right>=\left<\frac{1}{\gamma_{\rm particle}(v)} \right>$ with $H_{\rm particle}$ being the Hamilton function of an individual particle says how relativistic is the motion of the system. In the rest frame $K_0$ from (\ref{5.11b}) we can reconstruct both: nonrelativistic
$T_0= \frac{E_0^{\rm nonrel.}}{\frac{3}{2}Nk}$ and the ultrarelativistic  $T_0= \frac{\sum_{i=1}^N \sqrt{\vec{p}_i^{\,2} c^2}}{3Nk}$ limits. Notice that there is some freedom of the choice of the factor in Def. (\ref{5.11b}). Instead of 
$\frac{1}{3Nk}$ we can put $\frac{(\gamma(v))^r}{3Nk}.$ Thus for $r=1$ the statistical temperature is an invariant and for $r=2$ we obtain the transformation rule (\ref{2.27}).

Another possibility is suggested by the general formula (\ref{5.8}). First, we recall that the pressure $p$ is defined as the average momentum transported per second per unit area through the surface element in the direction given by the unit positive normal $\vec n$ to this element. It can be easily shown that using the above definition one gets \cite{rkp2}
\be
\label{5.19}
p=p_0=\frac{N}{V} \Big< (\vec{p} \cdot \vec n )[(\vec w -\vec v)\cdot \vec{n}]\Big>= 
\frac{N}{V} \Big< m \gamma(w)(\vec{w} \cdot \vec n )[(\vec w -\vec v)\cdot \vec{n}]\Big>,
\ee
where
$\vec w$ is the velocity of the particle with respect to the frame $K$, $\vec p$ as before is the particle momentum in the system $K$ and $\vec v$ denotes the velocity of the vessel containing gas relative to $K.$

From (\ref{2.27}), (\ref{5.8}) and (\ref{5.19}), remembering that $V=(\gamma(v))^{-1}V_0$ we find
\be
\label{5.20}
T= (\gamma(v))^2 \frac{1}{k} \Big< m \gamma(w)(\vec{w} \cdot \vec n )[(\vec w -\vec v)\cdot \vec{n}]\Big>.
\ee

This formula shows that the absolute temperature $T$ can be measured by the  `statistical thermometer' defined by the right hand side of (\ref{5.20}). However, the term $(\gamma(v))^2$ appearing in (\ref{5.20}) is somewhat artificial and is taken into account only in order to reproduce the absolute temperature $T.$ Other terms $(\gamma(v))^r$ can also be used.

Thus one can find the temperature as a Lorentz invariant \cite{ptl2}, \cite{dc} 
\be
\label{5.21}
T^{\rm (L)}=T_0= \gamma(v) \frac{1}{k} \Big< m \gamma(w)(\vec{w} \cdot \vec n )[(\vec w -\vec v)\cdot \vec{n}]\Big>
\ee
or the temperature transforming according to (\ref{2.28}) \cite{cr} (moving system appears to be cooler)
\be
\label{5.22}
T^{\rm (Planck)}= \frac{1}{k} \Big< m \gamma(w)(\vec{w} \cdot \vec n )[(\vec w -\vec v)\cdot \vec{n}]\Big>.
\ee
Analogous arbitrariness appears when one looks for the temperature of a moving black body by employing the Planck distribution 
\cite{ptl5}, \cite{ptl6}, \cite{ssc}.

{\bf Acknowledgment} Authors wish to thank  Dr. A. Montakhab and Dr. G. L. Sewell for indicating some interesting papers related to the present article.

\end{document}